\documentclass[aps,twocolumn,groupedaddress,prl]{revtex4-1}
\usepackage{amsmath, amssymb, graphics, epsfig, bbm}
\usepackage[latin1]{inputenc} 

\usepackage{color}
\usepackage{times}
\newcommand{\be}{\begin{eqnarray}}
\newcommand{\ee}{\end{eqnarray}}

\newcommand{\bee}{\begin{equation}}
\newcommand{\eee}{\end{equation}}
\newcommand{\one}{\mathbbm{1}}
\newcommand{\Tr}[1]{\mathrm{Tr}\left[ {#1} \right]}
\newcommand{\pTr}[2]{\mathrm{Tr}_{#1}\left[ {#2} \right]}
\newcommand{\ket}[1]{\left|{#1}\right\rangle}
\newcommand{\bra}[1]{\left\langle{#1}\right|}

\newcommand{\ketbrad}[1]{\left|{#1}\rangle\!\langle{#1}\right|}

\newcommand{\qed}{\ensuremath{\hfill \Box}}
\newcommand{\ie}{\emph{i.e.~}}
\newcommand{\eg}{\emph{e.g.~}}

\newtheorem{theorem}{Theorem}
\newtheorem{definition}{Definition}

\begin{document}

\title{Wick's Theorem for matrix product states}

\author{R.\ H\"ubener$^1$, A.\ Mari$^{1,2,3}$, and J.\ Eisert$^1$}

\affiliation{$^1$ Dahlem Center for Complex Quantum Systems, Freie Universit\"at Berlin, 14195 Berlin, Germany}
\affiliation{$^2$ Institute of Physics and Astronomy, University of Potsdam, 14476 Potsdam, Germany}
\affiliation{$^3$ NEST, Scuola Normale Superiore and Istituto di Nanoscienze - CNR, 56126 Pisa, Italy}
\begin{abstract}
Matrix product states and their continuous analogues are variational classes of states that capture quantum many-body systems or quantum fields with low entanglement; they are at the basis of the density-matrix renormalization group method and continuous variants thereof. In this work we show that, generically, $N$-point functions of arbitrary operators in discrete and continuous translation invariant matrix product states are completely characterized by the corresponding two- and three-point functions. Aside from having important consequences for the structure of correlations in quantum states with low entanglement, this result provides a new way of reconstructing unknown states from correlation measurements, e.g., for one-dimensional continuous systems of cold atoms. We argue that such a relation of correlation functions may help in devising perturbative approaches to interacting theories.
\end{abstract}

\maketitle


Quantum states of many-body systems or fields are characterized by their $N$-point correlation functions. Unsurprisingly, given their central status in the respective theories, there are many ways in which such correlation functions can be book-kept in terms of as simple as possible mathematical objects. For instance, prominent perturbative methods for the description of interacting field theories make extensive use of the relation between high order and two-point correlators \cite{PeskinSchroederQED,IsserlisWick}. These methods, supported by Isserlis' or Wick's Theorem \cite{IsserlisWick}, give rise to a practical way of identifying the propagators as the basic objects for the description of the situation at hand, as well as an interpretation in terms of virtual processes.

In this work we show that, remarkably, generic translation invariant matrix product states \cite{mpstheory,DMRG,Area,fannes:1991a} and their continuous analogues, cMPS or holographic states \cite{F,F3} are completely characterized by their two- and three-point functions. These states comprise a variational state class that approximates states with limited spatial entanglement well---a ubiquitous property for good reasons \cite{DMRG,Area}---and are at the basis of the seminal density-matrix renormalization group (DMRG) method \cite{DMRG} and continuous versions thereof \cite{F}. What is more, in our approach states and corresponding operators can be constructed such that their $N$-point correlation functions are completely characterized by their correlators of up to arbitrary odd order. We do so by proposing an explicit construction procedure of how to reconstruct higher-order correlation functions from lower order ones. This insight has a number of interesting consequences.

To start with, a fruitful research program has emerged in recent years of revisiting  questions in many-body theory within the variational set of matrix product states, now seen as a ``theoretical laboratory''. This approach has the appealing feature that some links and statements that are in all generality too hard to capture analytically can be formulated in a completely rigorous fashion. In this mindset, complete classifications of quantum phases have been given, new instances of Lieb-Schultz-Mattis theorems proven, or phase transitions of arbitrary order identified \cite{MPSLab}. Our statement provides a new tool to grasp the structure of matrix product states and their analogues for quantum fields.

Our result also identifies matrix product states as a variational class that is similar to, but yet beyond, quasi-free approaches. This observation may be even more interesting in the light of the fact that it is not straightforward to construct natural classes generalizing Gaussian states: For example, it is known that any unitary evolution generated by quadratic polynomials in the canonical coordinates maps Gaussian states to Gaussian states. If one looks at the closure of the unitaries generated by the quadratic polynomials and a single further term, say, of third order, one does not arrive at a meaningful new variational class, but in fact generates a set dense in all unitaries \cite{Lloyd}.

More practically speaking, our result clearly opens up novel ways to think of reconstruction methods for quantum states. We show how one dimensional lattice states and states of continuous systems such as cold atoms on top of atom chips \cite{SchmiedmayerAD} can be reconstructed or approximated using low order correlation data only.

Finally, the most important implications may come from a fundamental insight into the inherent structural properties of correlations as such. Our result shows that many physically relevant states of limited entanglement are corresponding one-to-one to families of meromorphic functions with interdependent poles.

{\it Matrix product states.} The main theorem of this paper applies to \emph{generic} translation invariant (continuous) matrix product states in the thermodynamic limit. Let us define what this means and fix some basic notation. A discrete matrix product state vector of an $\nu$-partite spin system with periodic boundary conditions is given by
\begin{multline}
	\ket{\psi_{\text{MPS}}} = \sum_{s_{\nu}, \ldots ,s_1} \Tr{A^{(\nu)}[s_{\nu}] \ldots A^{(1)}[s_1]}
	\ket{s_{\nu}, \ldots ,s_1},
\end{multline}
where $A^{(i)}[s_i]\in \mathbbm{C}^{d\times d}$ for all $i$. In this work we will focus on the thermodynamic limit, \ie $\nu \rightarrow \infty$, and the translation invariant case, \ie $A^{(i)}[s]=A^{(j)}[s]$ for all $i,j$. The finite bond dimension $d$ will be arbitrary but fixed. In this setting, correlation functions of a set of operators $\{O_{j}\}$ labeled by an index $j$ and with support on (different) sites $i_k$ with $0 = i_1 < \ldots < i_N$ take the form
\begin{multline}\label{MPSexpectation}
	\langle O_{j_N}^{(i_N)} O_{j_{N-1}}^{(i_{N-1})} 
	\ldots O_{j_1}^{(i_1)} \rangle \\
	=  \Tr{M^{[j_N]} E^{i_N-i_{N-1}-1} M^{[j_{N-1}]} \ldots 	M^{[j_1]} E^{\infty}} =: C^{(N)}_{\mathbf{j}}(\mathbf{n}),
\end{multline}
with $M^{[j]} = \sum_{m,n} A^*[m] \otimes A[n] \bra{m} O_j \ket{n}$, the transfer matrix $E = \sum_s A^*[s] \otimes A[s]$, and $E^{\infty}:=\lim_{n \rightarrow \infty}E^n$, which exists when the state is normalized. The star indicates complex conjugation of the matrix elements. We have written the distances in a compact form as $\mathbf{n}=(i_2-i_1-1, \ldots , i_N-i_{N-1}-1) \in \mathbbm{Z}^{N-1}$ and summarized likewise $\mathbf{j}=(j_1, \ldots ,j_N)$. It is possible to consider finite dimensional and infinite dimensional local systems; in the latter case the matrices $A[s]$ have to be chosen such that the infinite sums converge. 

The expectation values are invariant under simultaneous conjugation of all $M^{[j]}$ and $E$ with some invertible matrix, making it possible to consider an equivalent formulation where $E$ is in its Jordan normal form (JNF), \ie $E\mapsto J(E)$. We call the MPS \emph{generic} if $J(E)$ has non-degenerate diagonal entries $\mu_1, \ldots , \mu_{d^2}$ and, moreover, if the largest absolute value occurs only once. We order the diagonal elements by their absolute value, in descending order. Note that in the thermodynamic limit normalization implies $|\mu_i| \leq 1$, where the one with the largest magnitude equals unity, \ie $\mu_1 = 1$. In the future, we will simply say eigenvectors when we mean the \emph{right} eigenvectors, \ie $E \ket{i} = \mu_i \ket{i}$. The number of Schmidt coefficients, and hence the entanglement belonging to any contiguous bipartition of regions, is limited by $2d$.

{\it Continuous MPS.}
A one dimensional non-relativistic bosonic quantum field can be described in terms of field operators $\Psi(x)$ and $\Psi^\dag(x)$, with $[\Psi(x),\Psi(x')^\dag]=\delta(x-x')$ and $\Psi(x)|0\rangle =0$, where $|0\rangle$ is the vacuum. A particular class of one dimensional quantum fields is that of continuous MPS (cMPS) or holographic states \cite{F} with state vectors
\bee
	\ket{\psi_{\text{cMPS}}} = \pTr{\rm aux}{\mathcal P e^{\int_0^L dx Q(x)\otimes \one + R(x) \otimes \Psi^\dag(x)} }\ket{\Omega},
\eee
where $Q(x)$ and $R(x)$ are $x$-dependent finite-dimensional complex matrices acting in a $d$-dimensional auxiliary space. Similar to the case of MPS, we focus on translation invariant cMPS, having constant $Q$ and $R$, in the thermodynamic limit $L\rightarrow \infty$. It is useful to introduce the Liouvillian matrix 
\bee\label{defTmatrix}
	T = Q^* \otimes \mathbbm 1 + \mathbbm 1 \otimes Q + R^* \otimes R.
\eee
A state of such a quantum field is completely characterized by all the possible normal ordered correlation functions of the operators $\Psi(x)$ and $\Psi^\dag(x)$ \eg
\begin{equation}\label{corr}
	\langle \Psi^\dag(x_2) \Psi^\dag(x_5) \ldots \Psi(x_4) \Psi(x_3) \Psi(0)\rangle,
\end{equation}
where the order of position labels is such that they increase in size from left to right within the $\Psi^{\dagger}$, decrease within the $\Psi$, and $0 = x_1 < \ldots < x_N$. Correlation functions of cMPS are given by expressions involving only the auxiliary space. Let $e^{T\infty}$ be a short notation for $\lim_{L \rightarrow \infty}e^{T L}$; this limit makes sense when the state is normalized.

For translation invariant cMPS, we consider the differences between points, $\tau_i = x_{i+1} - x_i$ and summarize them in a vector notation $\boldsymbol{\tau}=(\tau_1,\tau_2, \ldots ,\tau_{N-1}) \in \mathbb R^{N-1}$. Let the matrices $M^{[j]}$ be equal to $R^* \otimes \one$, $\one \otimes R$ or $R^* \otimes R$ etc. With this notation we represent all $N-$th order correlation functions in a compact and straightforward way. For example 
\begin{multline}\label{cMPSexpectation}
	\langle \Psi^\dag(x_2) \Psi^\dag(x_3) \Psi(x_2) \Psi(0) \rangle\\
	=\Tr{M^{[1]} e^{T \tau_2} M^{[3]}  e^{T \tau_1} M^{[2]} e^{T\infty}} =: \mathcal{C}^{(3)}_{\mathbf{j}}(\boldsymbol{\tau})
\end{multline}
with $\boldsymbol{\tau}=(x_2, x_3 - x_2)$, $\mathbf{j}=(1,3,2)$, and $M^{[1]} = R^* \otimes \one$, $M^{[2]} = \one \otimes R$ and $M^{[3]} = R^* \otimes R$. Note that also in this case a gauge transformation is possible, corresponding to a simultaneous conjugation of $T$ and the matrices $M^{[j]}$ by an invertible matrix, so that we can always go to a picture where $T$ is in its JNF. The relationship between cMPS and channels directly implies \cite{channeltheory} that the diagonal elements $\lambda_1, \lambda_2,\ldots ,\lambda_{d^2}$ of $J(T)$ are closed under conjugation. We call the cMPS \emph{generic} if $J(T)$ has a non-degenerate diagonal and, moreover, the largest real part occurs only once. We order the eigenvalues in descending order by their real parts; the normalization of cMPS implies that it is non-positive and the largest one is zero, \ie $\lambda_1 = 0$.


{\it Main result.}
In general, to characterize the full state of a quantum system one needs to specify all the correlation functions. One may ask the following question: {\it ``Is it possible to completely characterize a (continuous) matrix product state from low order correlation functions?''} With the only initial assumption of a bond dimension $d$, we will show how to:
\begin{enumerate}
\item Certify that the given (c)MPS is generic. \label{oneA}
\item Reconstruct the full state of a (c)MPS from low order correlation functions once \ref{oneA}.\ has been verified.
\end{enumerate}
Both aspects will be studied in detail in the following.


{\it Data structure and transformations.}
We will use both the Z- and the Laplace transform of correlation functions in their multi-dimensional form. For discrete MPS, the Z-transform
\bee\label{zdef}
\mathcal{Z}^{(N)}_{\mathbf{j}}(\mathbf{s}) = \sum_{n_1, \ldots ,n_{N}} s_1^{n_1} \ldots s_N^{n_N} C^{(N)}_{\mathbf{j}}(\mathbf{n}),
\quad s_1,\dots, s_N \in \mathbbm{C}
\eee
is applicable.
Similarly, we have a Laplace transformation of the cMPS correlation functions
\bee\label{laplacedef}
\mathcal{L}_{\mathbf{j}}^{(N)}(\mathbf{s})=\int_0^\infty d^{N-1} \boldsymbol{\tau} e^{-\mathbf{s} \cdot \boldsymbol{\tau}} \mathcal{C}^{(N)}_{\mathbf{j}}(\boldsymbol{\tau}),\quad s_1,\dots, s_N \in \mathbbm{C}.
\eee
Depending on the correlation data, these transformations will not converge everywhere. The key observation is that under the assumption of a non-degenerate diagonal of $J(E)$, we have
\begin{multline}\label{strucmps}
C^{(N)}_{\mathbf{j}}(\mathbf{n}) = \sum_{k_1,\ldots,k_{N-1}=1}^{d^2} c^{(N)}_{\mathbf{j}}(k_1, \ldots ,k_{N-1}) \\
\times (\mu_{k_{N-1}})^{n_{N-1}} \ldots (\mu_{k_1})^{n_1},
\end{multline}
where
\begin{multline}\label{coeff}
c^{(N)}_{\mathbf{j}}(k_1, \ldots ,k_{N-1}) =\\ \bra{1} M^{[j_{N}]}\ketbrad{k_{N-1}} M^{[j_{N-1}]} \ldots \ketbrad{k_1} M^{[j_1]} \ket{1},
\end{multline}
and $\{\ket{k}\}$ denotes the basis where $E$ obtains JNF. Note that $E^\infty= \ketbrad{1}$. Considering $|\mu_i| \leq 1$ with $\mu_1=1$, we deduce that the Z-transform---as a function in the complex variables $\{s_1, \ldots, s_{N-1}\}$---converges within the product of unit disks around the origin. It is possible, starting from this region, to reconstruct the whole meromorphic function (its poles and the residues) by analytic continuation. Another way of dealing with \eg experimental data, would be to fit functions of the given form using only the unit disk as support. Summarizing, we have access to
\bee\label{ztrafofinal}
\mathcal{Z}^{(N)}_{\mathbf{j}}(\mathbf{s}) = \sum_{k_1,\ldots,k_{N-1}}^{d^2} \frac{c^{(N)}_{\mathbf{j}}(k_1, \ldots ,k_{N-1}) }{(1 - \mu_{k_1} s_1) \cdots (1 - \mu_{k_{N-1}} s_{N-1})}.
\eee
Similar considerations apply to cMPS correlations, which, under the assumption of non-degenerate $J(T)$, turn into
\begin{multline}\label{strucCmps}
	\mathcal{C}^{(N)}_{\mathbf{j}}(\boldsymbol{\tau}) = 
	\sum_{k_1,\ldots,k_{N-1}=1}^{d^2} c^{(N)}_{\mathbf{j}}(k_1, \ldots ,k_{N-1})\\ 
	\times	e^{\lambda_{k_1} \tau_1} \ldots e^{\lambda_{k_{N-1}} \tau_{N-1}}
\end{multline}
with the same symbols $c^{(N)}_{\mathbf{j}}(k_1, \ldots ,k_{N-1})$ as defined in Eq.\ \eqref{coeff}. The integral in Eq.\ \eqref{laplacedef} yields meromorphic functions in higher dimensions of the form
\bee\label{laplacetrafofinal}
\mathcal{L}^{(N)}_{\mathbf{j}}(\mathbf{s}) = \sum_{k_1,\ldots,k_{N-1}}^{d^2} \frac{c^{(N)}_{\mathbf{j}}(k_1, \ldots ,k_{N-1})} {(\lambda_{k_1} - s_1) \cdots (\lambda_{k_{N-1}} - s_{N-1})}.
\eee
Since $\mathfrak{Re} \lambda_i \leq 0$, the region of convergence of the integral in Eq.\ \eqref{laplacedef} is the product of complex half-planes with positive real part. Being a meromorphic function, it can be reconstructed using this data. Note that now $e^{T \infty} = \ketbrad{1}$. 
The fact that the diagonal of $J(E)$ and $J(T)$ is non-degenerate and has only finitely many elements played a crucial role in the derivation of the form of Eqs.\ \eqref{strucmps} and \eqref{strucCmps}.

{\it Reconstruction theorem.}
The form of the equations \eqref{ztrafofinal} and \eqref{laplacetrafofinal} implies that all poles are elements of $\{\lambda_i\}$ and $\{\mu_i^{-1}\}$, respectively. Depending on whether the corresponding residues $c^{(N)}_{\mathbf{j}}(k_1, \ldots ,k_{N-1})$ are zero or not, the transforms of the correlation data may or may not reveal poles at these points. This makes it useful to give the following definition.
\begin{definition}\textbf{\emph{($p$-number)}}
Given a (c)MPS with bond dimension $d$, we define the $p$-number as the minimum order $p$ such that $d^2$ distinct poles appear in the Z- or Laplace transforms, respectively, in at least one correlation function of order less or equal to $p$. If the minimum does not exist we say that the $p$-number is infinite.
\end{definition}
Note that in this definition, we only need \emph{one} correlation function of \emph{any} subset of operators of interest to show all poles in one of its arguments, in order to derive the $p$-number. This provides a solution to the first task: If the $p$-number of a (c)MPS is finite, we can directly claim that $E$ or $T$ have a non-degenerate Jordan diagonal.
Now we are going to study in more detail not only the poles but the full structure of correlation functions of (c)MPS. We work in the basis where $E$ or $T$ are in JNF. If a function $\mathcal{Z}^{(N)}_{\mathbf{j}}$ or $\mathcal{L}^{(N)}_{\mathbf{j}}$ is given, each coefficient $c^{(N)}_{\mathbf{j}}(k_1, \ldots ,k_{N-1})$ can be extracted by finding the residue of the corresponding multi-pole. We can finally state the main theorem:
\begin{theorem}\textbf{\emph{(Computing higher from lower correlation functions)}}
A generic, translation invariant (c)MPS in the thermodynamic limit with $p$-number $p$ is completely characterized by the correlation functions of order $\ell\le 2 p -1$.
\end{theorem}
\emph{Proof:} We merely need to consider the case in which $p$ is finite. From the definition of the $p$-number we know that $J(E)$, respectively $J(T)$, have a non-degenerate diagonal in the JNF, whose entries can be recovered from (the pole structure of the transforms of) the correlation functions of order $N \leq p$. This reduces the reconstruction of correlation functions to the reconstruction of the coefficients in Eq.\ \eqref{coeff}. 
The problem is now to express every coefficient
\be\label{coeff2}
c^{(N)}_{\mathbf{j}}(k_1, \ldots ,k_{N-1}) = M^{[j_{N}]}_{1,k_{N-1}} M^{[j_{N-1}]}_{k_{N-1},k_{N-2}} \ldots M^{[j_1]}_{k_1,1},
\ee
each associated with a unique set of poles, in terms of low order coefficients $c^{(\ell)}$ with $\ell\le 2p-1$. From the definition of the $p$-number we know that for each index $k$ there is at least one non-zero coefficient $c^{(p(k))}_{\mathbf{j}(k)}(...,k,...) \neq 0$, with $p(k) \le p$ and fixed $\mathbf{j}(k)$, having $k$ as one of its indices. This allows us to write $d^2$ different versions of the identity (this is a scalar)
\bee \label{partitionofunity}
1^{(k)}= \frac{M^{[j(k)_{p(k)}]}_{1,\star} \, \dots \,  M^{[j(k)_{\ell'}]}_{\star,k} M^{[j(k)_{\ell'-1}]}_{k,\star}  
\, \dots \, M^{[j(k)_1]}_{\star,1}}
{c^{(p(k))}_{\mathbf{j}(k)}(...,k,...)},
\eee
where $k=1,2,...,d^2$, and the symbol $\star$ stands for other indices which are irrelevant. Now we reorder the matrix elements in the numerator by shifting all matrix elements on the r.h.s.\ of the index $k$ simultaneously to the l.h.s., leaving the order of the other indices untouched, \ie in the following way
\bee\label{onek}
1^{(k)}= \frac{M^{[j(k)_{\ell'-1}]}_{k,\star}  
\, \dots \,M^{[j(k)_1]}_{\star,1} M^{[j(k)_{p(k)}]}_{1,\star} \, \dots \,  M^{[j(k)_{\ell'}]}_{\star,k}}
{c^{(p(k))}_{\mathbf{j}(k)}(...,k,...)}. 
\eee
We can finally put all these resolutions of the identity between the matrix elements in Eq.\ \eqref{coeff2}
\begin{multline}\label{residues}
c^{(N)}_{\mathbf{j}}(k_1, \ldots ,k_{N-1}) = \\
M^{[j_{N}]}_{1,k_{N-1}} 1^{(k_{N-1})} M^{[j_{N-1}]}_{k_{N-1},k_{N-2}} 1^{(k_{N-2})} \ldots 1^{(k_1)} M^{[j_1]}_{k_1,1}.
\end{multline}
We recognize, in the numerator, several new strings of matrix elements resulting from the insertion. They have the same structure as in Eq.\ \eqref{coeff2} but a lower order $\ell\le 2p-1$. This means that, for every $N$, all the coefficients $c^{(N)}$ can be written in terms of $c^{(\ell)}$ with $\ell \le 2p-1$. In other words, correlation functions of order less or equal to $2p-1$ are enough to reconstruct all the others. This proves the validity of the theorem.\qed

{\it Example.}
It is instructive to consider the following case. Given an MPS or a cMPS with finite $d$, let the operators $O_j$ and the state be such that the corresponding matrices $\tilde M^{[j]} = X M^{[j]} X^{-1}$ have only \emph{non-zero} elements. Here $J(\cdot)= X \cdot X^{-1}$ is the conjugation that takes $E,T$ to their JNF. Note that the probability to have this situation in an experiment, or using a randomized (c)MPS and operators $O_j$, is one. Under this condition, all two-point function transforms show all the poles, hence $p=1$. Computationally, all residues of all the poles of all $N$-point functions with $N \leq 3$ can be obtained. Hence we can, using the construction above, give explicit formulas that express all $N$-point functions in terms of the $2$- and $3$-point functions. 

{\it Applications in tomography.}
The framework established here opens up novel ways to reconstruct unknown low-entanglement states from correlation data alone. Our approach gives rise to a complementary picture to the method of reference \cite{mpstomo}, where the reconstruction is based on tomographic estimation of certain reduced states. We moreover address quantum field states and continuous systems, for which no method is known altogether. Consider the application to correlation data of atom counting experiments, or of split and recombined Bose condensates \cite{SchmiedmayerAD,Errors}. See the Supplementary material for detailed instructions.

{\it Location of poles and decay behavior.}
The decay behavior of contributions to the correlations follows directly from the position of the poles on the complex plane: note the relation between poles and diagonal elements: $s_i = \mu_i^{-1}$ and $s_i = \lambda_i$, respectively. The MPS poles describing slow decay are sitting close to the unit circle, and the cMPS poles of this kind are close to the imaginary axis. An inclusion of matrix dimensions in a reconstruction of only a subspace of the auxilliary space can be guided by the relevance of the poles for the desired range of correlations.

{\it Outlook.}
A stimulating insight is given by the mathematical structure of the correlations, which are, as shown, related to meromorphic functions with interdependent residues. The structure of correlations is moreover linked to the quantitative limitation of entanglement between spatial regions on a fundamental level. 

One might speculate that based on the findings above---in particular the relations of higher order correlations to two- and three-point functions---some new insight into diagrammatic perturbative methods could be obtained. One future direction of investigation are virtual processes. In well-known perturbative methods for interacting field theories, the states of the interacting theory are described in terms of states of the non-interacting counterpart. This leads to a description in terms of quasi-free states, determined by their two-point correlators, and puts the focus on propagators and an interpretation of the theory in terms of virtual processes. While this approach allows to predict experimental data with very high precision \cite{PeskinSchroederQED}, it gives rise to conceptual problems \cite{Haag}. Hence, transfer of the structures discussed here to a relativistic setting might be interesting. 

Of course, the relations underlying our approach are already summarized in a computationally efficient formal framework: the family of states known under the name of MPS, cMPS, and instances of projected entangled pair states. This means, given experimental data with the relations above, an optimal book-keeping device would be \eg an MPS. However, say, series expansions in a perturbation approach starting from MPS \emph{and potentially leaving this class of states} due to closing gaps etc.\ give rise to different sets of terms with different structure, and the MPS scheme is only one possible way to interpret it. A different summation order might yield a different optimal book-keeping device in this context.

In a related line of thinking, there is the possibility that the meromorphic structure of the correlations, together with the interdependencies of the poles, enable a different mathematical understanding of the underlying renormalization procedure. Such an understanding might help to find variations of the renormalization procedure, including possibly meaningful and computationally efficient extensions of MPS and cMPS to higher dimensions.


{\it Acknowledgements.}
We would like to thank the EU (Qessence), the EURYI, the ERC, and the BMBF (QuOReP) for support. We warmly thank Tobias J.\ Osborne for discussions, who emphasized that the formalism developed	originally for cMPS applies equally well to MPS, as well as M.\ Cramer for comments on tomographic issues.


\section{Appendix}

\subsection{Quantum dynamical semi-groups}

There is a strong formal relationship between cMPS with bond dimension $d$ and quantum channels which are elements of a  completely positive continuous one-parameter semi-group in $d$-dimensional Hilbert spaces, called Markovian quantum channels. For such elements $T_x$ there exists a generator ${\cal L}$ with ${\cal L}^\ast (\one)=0$ such that
\bee
	T_x= e^{x {\cal L}}
\eee
for all $x\geq 0$. The generator ${\cal L}$ of the family of quantum channels 
is necessarily of the form
\bee
	{\cal L}(\rho) = - i[\rho,H] - \frac{1}{2}
	\sum_{j} \left(R_j^\dagger R_j \rho + \rho R_j^\dagger R_j  - 2 R_j \rho R_j^\dagger
	\right).
\eee
Using the isomorphism relating a density matrix defined on a $d$-dimensional Hilbert space to its vector form in a $d\times d$-dimensional Hilbert space, 
\bee
	\rho(x) \mapsto | \rho (x) \rangle = \sum_{i,j=1}^{d} \rho_{i,j}(x)| i \rangle |j \rangle,
\eee
the above generator takes the form
\be
	L &=&  - i H^\ast \otimes \one + i \one\otimes H \nonumber\\
	&-& \frac{1}{2}
	\sum_{j} \left(
	R_j^T R_j^\ast \otimes \one + \one\otimes R_j^\dagger R_j - 2 R_j^\ast\otimes R_j
	\right).
\ee
The Liouvillians encountered above are Liouvillians in this sense, where we merely have a single Lindblad operator $R$. 
With 
\bee
	Q=iH - \frac{1}{2}
	R^\dagger R,
\eee
one has 
\be
	L &=&  Q^\ast \otimes \one +  \one\otimes Q  
	+ R^\ast\otimes R,
\ee
which the Liouvillian defined in the main text. 
Since the spectrum of ${\cal L}$ is taken to be non-degenerate, 
there exists a unique stationary state, given by 
\bee
	\rho_{\rm ss}= \lim_{x \rightarrow \infty} e^{x {\cal L} } (\rho) 
\eee
for every $\rho$. The generator is also gapped, so that convergence to the stationary state is exponential. Such families of quantum channels are sometimes referred to as being relaxing.

\subsection{Applications in tomography}\label{sec:app_tomo}

The framework established here opens up novel ways to reconstruct unknown low-entanglement states from correlation data alone, in the sense of tomographic reconstruction. Our ideas seem promising for, but are not limited to, the tomographic reconstruction of quantum field states and continuous systems, for which no method is known altogether. Consider the application to correlation data of atom counting experiments, or of split and recombined Bose condensates \cite{SchmiedmayerAD}, where a reconstruction of quantum field states is disirable and possible using our approach \cite{Errors}.

The result obtained here is primarily a structural statement: Surprisingly, MPS and cMPS are generically fully determined by their two- and three-point correlation functions. Now, considering that for a scalable procedure, one necessarily has to make use of data structures that are efficient in the system size, this structural properly is very useful for a reconstruction of states. Both for MPS and cMPS, this insight gives rise to a tomographic method. For the former, the tomographic procedure related to the structural analysis in the formal framework as given here is complementary---yet completely different in spirit---to the one recently given in Ref.\ \cite{mpstomo}. It is instructive to stress the differences between both methods.

In Ref.\ \cite{mpstomo} it had been considered how to reconstruct a generic MPS/MPO from the \emph{reduced density matrix} $\rho_k$, supported on $k$ contiguous sites, if a technical invertibility criterion is met. Upper bounds for $k$ can be deduced for a given bond dimension $d$ by applying the theory of Ref.\ \cite{fannes:1991a}. 
As a rule of thumb, one has to combine contiguous local sites such that the resulting effective local dimension is larger than or equal to the matrix dimension $d$. The number $k$ will hence usually be much larger than $3$, 
depending on the matrix dimension $d$. Eventually, the procedure in Ref.\ \cite{mpstomo} relies on complete tomographic knowledge of $\rho_k$ to reconstruct the MPS/MPO.

In this article, in contrast, we state that the MPS---as a functional over certain operator sets---can be reconstructed from the low-order \emph{correlation functions} of a fixed degree. In our case, the operator algebra does \emph{not} have to be tomographically complete. Moreover, in almost all cases, and essentially independent of the bond dimension $d$, the fixed low-order correlators mean two- and three-point functions (see $p$-number). This is a different result in an interesting way:
the knowledge of the three-point functions (which can be confined to finite distances) 
is opposed to the complete tomographic knowledge of a reduced density matrix of contiguous sites with $d$-dependent support.

Let us review the technical details of our reconstruction procedure. The aim of the reconstruction is to give a representation of a positive functional on a representation of operators. Hence, the quantities to be derived from the data are the matrices $J(E)$ and $M^{[j]}$, or $J(T)$ and $M^{[j]}$, respectively. Because the origin of the formulas of our reconstruction is a (continuous) MPS ansatz---modified by some gauges only---we are already certain of the positivity of such expressions. We consider two more aspects.

Firstly, any reconstruction procedure can at best find a representative of the 
equivalence class of all (continuous) matrix product states that are identical up to a gauge transformation. This is a feature and not a 
problem, since the state is clearly merely defined by this equivalence class.
From the above considerations, 
we immediately derive that simultaneous conjugations of the transfer matrix $E$ and the operator matrices $M^{[j]}$---analogously the Liouvillian $T$ and $M^{[j]}$---by an invertible matrix leaves the expectation values invariant. So does the transformation $Q \mapsto Q + \alpha \one$, which changes the norm of the cMPS. Some of these transformations have been performed to obtain the pole structure in the explicit form, diagonalizing the $E$ and $T$ and ordering the diagonal elements. Hence we have ``fixed the gauge", but only partially.

Secondly, we need a promise of the dimensionality $d$ of the matrices. The $p$-number then tells us if the experimental data is consistent with this promise and if we ``see enough structure in the data'', \ie if we can use it for a full reconstruction of one representative. 
As we have seen above,
in all cases except a set of measure zero the matrices $M^{[j]}$ contain no zeros as elements and the $p$-number is one. This is the only case to consider for practical applications, hence we will restrict the following discussion to this case. Let us moreover assume for simplicity to have only one operator $M$. The case of more $M^{[j]}$ is then straightforward. 

The reconstruction of the representative is then performed in the following way. One half of the task is to find the position of the poles $\{\mu_i^{-1}\}$ of the Z-transform of the correlators. Only the two-point function is needed to find the poles. The second half of the task is to find the residues $c^{(N)}_{\mathbf{j}}(k_1, \ldots ,k_{N-1})$, for $N=2,3$, which determine the operator matrices $M$. For this, we need the two- and three-point-function. For both tasks, we apply the Z-transform  to the data. Although the two-and three-point functions consists of an infinite amount of data points, only a small finite subset is required: the exponential convergence of the Z- transform allows an accurate approximate calculation with a finite number of data points (the actual number depending on the position of the poles). We hence essentially deal with sums of geometric series. Once we have the poles, we have $J(E)$. Regarding the corresponding residues, we consider the remaining unfixed gauge freedom of our representatives. If one uses a scalar $a_k \neq 0$
\be
M_{k,1} \mapsto a_k M_{k,1}, \quad M_{1,k} \mapsto a_k^{-1} M_{1,k}
\ee
then the $M$s change, but the two- and three-point functions (and the pole structure) do not. This gauge corresponds to the conjugation of $E$ and $M$ with an invertible diagonal matrix and has not been fixed before. We use this freedom to set $M_{1,k} = 1$. Accordingly, we obtain directly from the residues of the poles
\be
M_{k,1} = c^{(2)}(k).
\ee
The remaining elements of $M$ are obtained (see the general expressions in the main text) via
\be
M_{k,\ell} = c^{(3)}(k,\ell) / c^{(2)}(\ell).
\ee
This is sufficient for a reconstruction. If we have more data (\eg higher $N$-point functions), it can be used as a consistency criterion for the assumptions. If the data is noisy, the formulas will not be consistent, but maximum-likelihood and other estimators can be used and can profit from more data and the relations discussed in this article (compare also Refs.\ \cite{e1,e2,e3}).
This procedure is stable under the influence of noise in the sense that the residues of the poles and their positions are continuous in the matrix entries. For the purposes of this work, we will postpone questions of the scaling of the computational effort of the classical post-processing in the precision of the input \cite{Complex} and focus on the possibility of reconstruction of higher correlation functions from lower ones as such.


\begin{thebibliography}{99}
\bibitem{Errors}
	A detailed study of robustness of the scheme with respect to errors, compare Refs.\ \cite{e1,cs2}, will be presented elsewhere.
\bibitem{Complex}
	For a discussion of the computational complexity of deciding whether
	or not a family of completely positive maps is generated by an
	underlying Markovian master equation from data taken at finitely many
	times, see Ref.\ \cite{hard}.
\end{thebibliography}
\end{document}